# AI's Impact on Traditional Software Development: A Technical Look

Bhanuprakash Madupati

MNIT, MN

JULY 2024

**Abstract**
The application of artificial intelligence (AI) has brought key shifts in conventional tactical software development, including code generation, testing and debugging, and deployment. Waterfall and Agile development approaches, which have been used for a long time, also widely employ manual and well-planned steps. However, with the help of automated tools and models such as OpenAI Codex and GPT-4, many aspects of the Software Development Life Cycle (SDLC) have been made possible. This paper examines the technical aspect of integrating AI into prior traditional software development life cycle methodologies, emphasizing code automation, intelligent testing frameworks, AI-based debugging, and continuous integration and deployment pipelines. The analysis is also based on the advantages of utilizing AI for optimizations in efficiency, accuracy, and development speed alongside issues like over-dependence on AI, ethical questions, and technical constraints. Based on the case and example given in this paper, it is clearly shown that the self-improvement of AI in software development makes the process more dynamic, autonomous, and optimized.

## 1. Introduction

### 1.1 Background

The Waterfall, Agile, and DevOps models dominate the IT industry and have become the grounds for constructing advanced software solutions. These approaches follow a linear or circulatory model where coding, testing, and deployment steps are usually involved. Albeit efficient, these methods may have issues concerning timeliness, expansiveness, and errors due to human interference. With changing business environments and technologies, keeping pace and delivering the software in record time has become essential. As a result, the software development life cycle's (SDLC) dynamism has been recorded by incorporating Artificial Intelligence (AI).

Artificial intelligence has advanced much in integrating key segments of software processing, namely code generation, testing, debugging, and deployment. Through the use of OpenAI Codex and GPT-4, among other technological advancements, it is now possible to write functional code from simple text inputs, thereby greatly cutting down on the time required to code and minimizing human errors. Similarly, AI-assisted testing tools improve traditional testing techniques by detecting bugs and security breaches faster and more efficiently. AI in these areas enhances both efficiency and the quality and accuracy of software applications.

### 1.2 Purpose of the paper

Technically, this paper investigates how AI affects the conventional software development paradigms, specifically focusing on how AI tools and approaches disrupt several misery steps of code generation, testing, debugging, and deployment. The objectives of this paper are as follows: The objectives of this paper are as follows:

It is important to print a selection of the identified sources to understand better what areas of AI have impacted the original software development process.
With a primary emphasis on efficiency, accuracy, and cost, it's necessary to determine what value AI can contribute to SDP. To explain the main issues with AI applications in the SDLC process, including automation dependency and ethical problems. Thus, the paper offers an overview of AI's presence and potential in establishing the vision for software engineering.

**1.3 Scope and Limitations**

Specifically, while this paper will not concentrate on AI's social implications for the software industry, it will focus mainly on the technical side of the argument, exploring AI's effects on software writing, testing, debugging, and deployment. Although AI impacts diverse domains, this work will focus on concepts and technologies created and incorporated into the software sector up to March 2024. The AI and machine learning examples used in the analysis will be derived from large-scale and agile development environments.

**2. An insight into the SDLC – traditional Analysis of the Software Development lifecycle**
**2.1 Key Phases of Traditional SDLC**

Structured Software Development processes like SDLC have a very sequential structure and have long been divided into phases, such as the Planning, Design, Development, Testing, Deployment, and Maintenance phases. In conventional approaches, all stages predominantly rely on manual operations, and people's involvement is inevitable at most steps.
Planning and Requirement Analysis: In this phase, stakeholders help define the project's needs and expectations with the development team. This is usually a lengthy process since documentation and analysis are done thoroughly to satisfy all business requirements. Nevertheless, there is a lack of stewardship during this phase, where the stakeholders and developers acquire discrepancies, resulting in later revisions [4].

Design and Architecture: This phase involves developing the software structure and ensuring it meets the required needs. It is typically a manual phase where system architects design intricate strategies and architectures of the systems, utilizing various conceptual tools such as Unified Modeling Language diagrams. The outcome is a clear and concise map to follow through to the coding phase of the project.

Development (Coding): In the development phase, the program's design is converted into code that the computer can execute. As per the architecture diagrams, large-scale applications are developed by writing smaller programs to be integrated into languages such as Java, C++, or Python. This phase might be affected by human errors, which can result in the creation of some bugs or lower efficiency, which will be noticeable in future phases [2].
Testing: Testing in traditional SDLC is traditionally done after the developmental phase. This phase involves physical or automated testing to check whether the system performs well and as planned. However, although some level of automated testing is employed, most testing remains manual, contributing to time and effort in error detection and remediation [1].

Deployment and Maintenance: Similar to the other phases in traditional SDLC, the deployment phase is sequential, in which the software is rolled out in separate stages, and the upgrades and bug corrections are typically done by hand. There is frequent maintenance, correlation, and manual patching; updates are made to correct the bugs and enhance performance. [6]

**Table 1** *Overview of Traditional SDLC Phases and Key Challenges*

| Phase | Description | Key Challenges |
|---|---|---|
| **Planning and Requirement** | Gathering requirements and creating project documentation. | Risk of misalignment with stakeholders' expectations. |
| **Design and Architecture** | Developing system architecture and design using UML or similar. | Slow, rigid process; difficult to adapt to changes. |
| **Development (Coding)** | Manually code the software based on the design specifications. | Prone to human errors and delays. |
| **Testing** | Manual and semi-automated testing, focusing on bug identification. | Time-intensive and error-prone. |
| **Deployment and Maintenance** | Manual deployment and ongoing maintenance. | Sequential and time-consuming delays in fixing bugs. |

**2.2 Difficulties Of Traditional SDLC**

While traditional SDLC provides a structured approach to software development, it also comes with several challenges: While traditional SDLC delivers a structured approach to software development, it also comes with several challenges:
Slower Development Cycles: Traditional SDLC is sequential, which takes longer in the overall development process. Because each stage must be performed before the next stage, problems in one stage may influence all aspects of the project [6].

Difficulty Adapting to Changes: Non-flexible models such as the Waterfall do not cater to changes in requirements in midstream. Thus, the project may be suspended, or extensive modifications may be required in response to a change in business requirements—both expensive [4].
High Cost of Error Detection and Correction: Many errors are inserted in the early stages of development and are not easily identified until the testing or even the deployment phase, which is costly and time-consuming [1].

**Graph 1**: AI-Powered Testing Efficiency

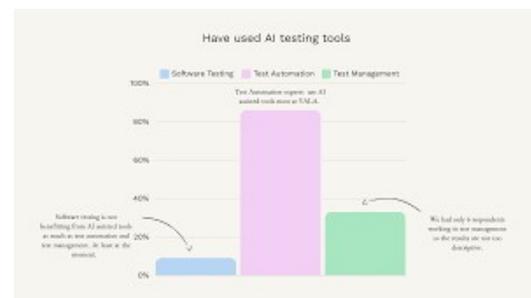

This is demonstrated in the following graph, where Test Automation employs approximately 80% AI, while Software Testing and Test Management employ only 20% and 40% AI, respectively.

**3. AI in Software Development**
**3.1 AI technology and its applications in the SE discipline**

AI is becoming prominent in software development by replacing traditional human interventions with automation, enhancing work productivity, and minimizing errors. Three key forms of AI responsible for this transformation include Machine Learning, Natural Language Processing, and Reinforcement Learning.

Machine Learning (ML): Deep learning models are widely applied to improve many aspects of the software development process, including bug discovery and code enhancement. These algorithms identify various patterns within the code, format it, correct it, and even forecast other glitches, making the code cleaner and more efficient [1], [4]. Also, ML plays a pivotal role in testing in which models estimate which parts of the codebase are more likely to contain bugs with the help of historical data to optimize testing procedures [1].

Natural Language Processing (NLP): NLP also involves code generation using languages such as OpenAI Codex and GPT-4. They translate instructions in natural language into code and have become useful in developing tools to create executable code snippets. This results in a considerable decrease in the number of hours devoted to completing sets of codes and allows for an increase in development speed [6]. Using NLP tools is highly beneficial for rapidly prototyping and cutting the development cycle short [1]. Reinforcement Learning (RL): RL improves decision-making during software testing by focusing on dynamic testing and debugging. As a result, RL models can adapt the test cases according to feedback received during software execution, and these cases may cover important scenarios that are often overlooked in other testing methods [1].

Graph 2. AI in software development

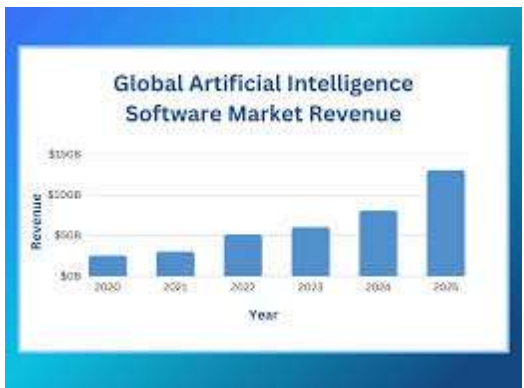

### 3.2 AI tools in software development

AI integration has revolutionized software development through automated code generation, testing, and debugging, previously manual processes. These tools facilitate rapid feedback cycles, minimize the occurrence of errors, and enhance development efficiency throughout the process.

**Table:** AI Tools in Software Development

| AI Tool | Application | Impact |
| --- | --- | --- |
| OpenAI Codex, GPT-4 | Automated code generation from natural language | Reduces manual coding efforts, accelerates development speed, and improves code quality by minimizing human error. |
| AI Testing Frameworks | Automated generation and execution of test cases | It improves test coverage, detects bugs earlier, and reduces the manual effort required for testing. |
| AI Debugging Tools | Predictive error detection and automatic bug fixing | Faster bug identification, reduced debugging time, and enhanced code reliability through proactive fixes. |

| CI/CD Automation Tools | Automates Continuous Integration/Deployment pipelines | It accelerates deployment cycles, reduces manual interventions, and improves system stability with automated rollbacks. |

AI-Assisted Code Generation: Current AI models such as OpenAI Codex and GPT-4 have revolutionized how developers code. These models can understand natural language inputs and generate the code themselves, which saves a lot of time. Aside from the general effects of increasing efficiency and speeding up development, AI saves time and reduces errors in coding. These Tools are efficient in large projects, especially when data quality and consistency are issues [6].
Automated Testing with AI: Computer-aided testing has revolutionized software testing. These tools use ML algorithms to forecast where bugs are more probable to exist, produce test cases, and run the tests themselves. This automation helps lessen testing efforts and detect key bugs at an earlier stage [1]. Furthermore, these tools enhance regression testing efficiency since it is time-consuming and resource-intensive in large-scale software development projects [1].

Fig 1: Ai software development process

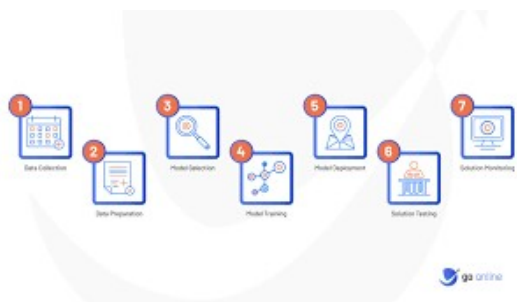

## 4. Main Technical Effects of AI on Conventional Software Development
### 4.1 Code Generation

AI greatly revolutionized traditional coding by automating most parts of the coding process. Applications such as OpenAI Codex and GPT-4 can take plain English instructions and write corresponding code. These models minimize coding by performing repetitive efforts, reviewing templates for coding, and automating various code works. Therefore, developers can concentrate on more significant problems like choosing the architecture and fine-tuning instead of typing codes [6].
AI also plays an important part when it comes to the enhancement of code quality. Coding is another area where AI can supercharge the programming process by using data sets containing millions of prior coding patterns and can generate flawless and efficient code with low chances of bugs and errors. This is especially important in extensive software development programs, where cohesion and productivity are critical [6]. When code generation is incorporated with AI, the development rates are quicker, and less time is spent mitigating problems at a later stage [6].

### 4.2 Testing and Quality Assurance (QA)
Testing and QA are considered to be essential stages of SDLC, and earlier, only manual or partially automated testing techniques were used. There is a noticeable efficiency enhancement due to AI automation of most testing procedures. Intelligent testing scripts can write tests, execute them, and even identify regions that may potentially contain faults [1]. This approach helps to save a lot of time than would be used in testing manually and covers many test cases than would be possible when testing manually [1]. Besides, machine learning models are often employed in regression testing, which involves using algorithms to determine which parts of the software are most likely to fail following modification. In terms of regression testing, which is often performed by developers who want to ensure that new changes do not affect previous code, AI shortens the time spent on this process [1]. This makes it possible to release more system updates frequently without compromising the integrity of the software [1].

**4.3 Debugging and Bug Fixing**

Debugging can also be improved through the enhancement of tools that can independently identify and correct errors in the code brought about by AI. Conventional debugging paradigms are highly manual, and programmers spend considerable time trying to isolate and fix bugs. Automated tools, on the other hand, use machine learning and predictive analytics to search for bugs in code repositories and suggest solutions in real time [1]. These tools minimize the time taken for debugging and hence boost total productivity, enhancing the development cycle [1].

In addition, predictive models can also identify troublesome areas mainly because of the patterns identified from previous completed projects. This allows developers to prevent issues that may cause more significant problems in the application, as they can detect them before they become severe errors [1].Integrating AI in the debugging process increases the code's reliability levels and reduces the number of post-deployment bugs, enhancing the resilience of software systems [1].

**Figure 4:** Bug fixing in AI-driven verification

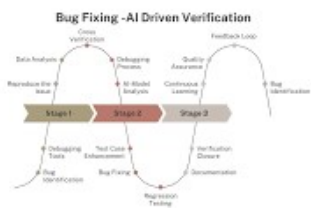

4.4 CI/CD Pipelines

In CI/CD, which stands for Continuous Integration/Continuous Deployment, AI intervenes by carrying out different steps within a pipeline. Conventional CI/CD processes entail the supervision and management of the build, testing, and deployment of code changes, which are, at times, tedious and vulnerable to human interventions. AI, on the other hand, speeds up this process of building, testing, and deploying models by minimizing human interference [7].

**Figure 4:** AI in CI/CD Pipelines

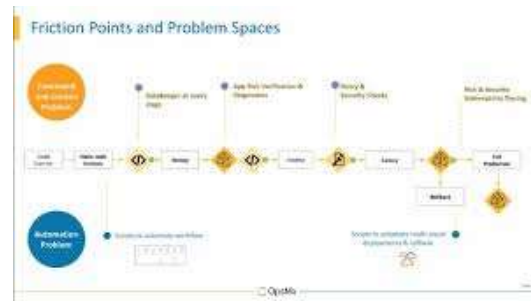

The AI models are constantly watching the pipeline; deployments are done automatically whenever the system notices that there are stable builds, and even faulty deployments can be rolled back with feedback from the system itself [7].This results in shorter cycles between builds and greatly minimizes the disruption of operations during the deployment of a new build, giving the development cycle more fluidity and the ability to adapt to changes. Through such automation, AI frees up the development teams and enables them to deliver new features and updates more regularly without compromising the quality [7].

**5.Case Studies and Applications**

**5.1 Overview of Use of AI in Large-Scale Projects**

AI has been implemented in several large projects, and it has drastically improved the software development processes.

Facebook's AI-Enhanced Testing Frameworks: Facebook has incorporated testing AI to help identify bugs as well as increase the coverage of tests. With the help of machine learning algorithms, Facebook can analyze which part of the code is most prone to failures during updates; therefore, the team can spend more time testing risky sections of the code. This strategy has significantly decreased the time spent on traditional testing and enhanced the quality of Facebook's new releases [1].

Google's AI-Driven Code Optimization in Cloud Applications: Auto-optimization: Google has incorporated AI to identify expensive code within its cloud applications and provide suggestions on how to improve it. AI models analyze the code in real-time to identify suboptimal performance and possible enhancements, which helps Google meet cloud services performance standards [7]. Through AI, Google's teams can concentrate on innovation because efficiently tested and improved code results from the process [7].

**Fig 5:** Testing in AI application

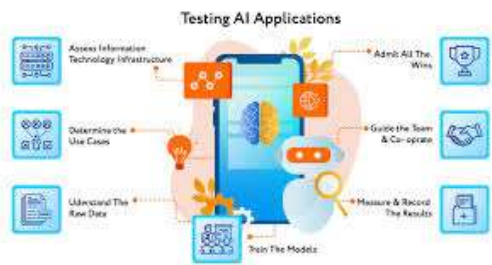

**5.2 AI in Agile Development**

It is also finding its way into Agile development environments characterized by high levels of iteration and the ability to adapt.Another critical feature of agile development involves delivering feedback and integrating it in the system at the highest possible rate, and this is where AI shines by generating code automatically and testing it [7].

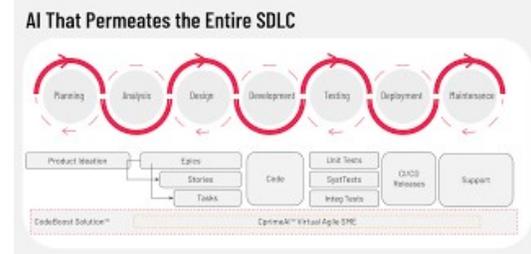

In Agile cycles, AI-driven tools like Codex and GPT-4 come into play and act as developers by writing code based on user stories and requirements. This improves the cycle time, enhancing teams' ability to work in parallel and offer new features at a faster rate [6].Also, AI-generated testing software also takes the responsibility of forming test cases for every new element brought in the Agile model as a feature, which means that quality will not be traded for speed [1].

The probability AI provides the identification of potential bottlenecks in sprints additionally benefits Agile procedures by enabling project managers to organize resources and development workloads efficiently.This guarantees that Agile iterations are on schedule while the chances of an unpredicted holdup are greatly mitigated [7].

**6.: Implications and Threats of AI Implementation in Software Development**

**6.1 Over-Reliance on AI Systems**

Another major concern that has been voiced regarding the rising use of Artificial Intelligence in software development is the likelihood of automation of some of the most basic and crucial duties in software development and management using AI.Although AI can bring many benefits when applied to the SDLC by automating and optimizing processes, a reliance on AI tools may lead to developers becoming less disciplined when working with their tools.Such dependency may lead to a complete removal of the human-in-the-loop, where important decisions are left to the AI models, thus, the likelihood of coming up with mistakes that might not be realized until some advanced stage of the development phase [6], [7] is highly likely.

Dangers of Excessive Dependence: Issues of complacency arise when developers rely heavily on AI.If most of the AI tools perform most of the coding and testing, then the developers might become disoriented with the structures and lack proper knowledge about the software.This becomes a critical issue in cases where an AI-generated code might contain seemingly minor defects or sub-optimal parts that are not easily discernible by the human brains [6].

**6.2 Ethical and Bias Concerns**

Ethical issues arising from the adoption of AI-driven software development are as follows: Ethical issues associated with bias in AI models and the impact of such biased models on society. Machine learning algorithms trained on data that contain bias are likely to have the same bias reflected in the code, ending up with biased code recommendations or testing practices. Such biases can also show up in the output of software that may create or support processes that result in unfair or discriminatory end-user consequences [8], [7].

Bias in AI-Generated Code: Literally, AI models are trained with datasets; they learn; if these datasets are biased then the bias will be reflected in the code that's written.This is especially the case with applications where fairness and equity are paramount, like those in the areas of medical diagnoses or the administration of justice [6].Reducing bias entails the constant assessment of the training datasets used for developing AI models and the implementation of fairness tests into software development processes.

Ensuring Fairness and Transparency: This specifically brings the issue of transparency as one of the vital factors that need to be considered while trying to avoid biases during the development of AI models.It is mandatory for the developers to make sure that the processes, underlying the code generation and the test automation are comprehensible, reproducible, and easily auditable.This level of transparency is required to ensure that people have trust in AI systems especially in situations where software is used to control essential infrastructures or even personal information.

**6.3 Technical Challenges**

Incorporating AI into traditional SW development methodologies also allows for the identification of several technical issues. AI tools may not be integrated with existing systems, especially where older systems are used in production lines.Also, AI models are unable to meet special organizational requirements that entail a great deal of expertise in using software [4], [6].

Compatibility Issues with Legacy Systems: In many industries today there are business applications and IT systems that are at least 10-20 years old and still running.AI models are most often designed with the use of current systems in mind and may pose issues when integrated into these systems.AI integration into such landscapes is often manual and time-consuming, which defeats the purpose of automation [4].

Inaccuracies in Specific Contexts: Techniques, especially machine-learning models developed using general datasets, will likely fail in narrow or specialized industries. For instance, data sources or systems involved in healthcare or finance have different regulatory constraints and complex structures compared to other sectors that may create chaos to the generalized AI models in coding, testing or deployment solutions [6].These challenges call for industry-specific datasets and creation of new AI models [4], [6].

**7.Future Prospects**

**7.1 Future of AI in Software Development**

As AI progresses in the future, the importance of AI in software development will increase significantly as it forms the basis for further development and will change the processes of creating and maintaining software.Therefore, AI is expected to be a more prominent part of the software lifecycle in the future with more tasks of the software lifecycle, such as integration, deployment, testing or maintenance, being fully or partially automated by 2030.

Artificial Intelligence Transforming Development by 2030: While current uses of AI in software development involve mostly code creation and bug identification, these capabilities will progress to encompass additional elements of software design and project organization.This could lead to the creation of self-evolvable systems, in which software programs are capable of evolving themselves depending on predefined parameters such as performance, and feedback from users.

Emerging Technologies: New technologies like quantum computing and 5G will shape the future of AI-driven software development. Of these emerging technologies, quantum computing, in particular, has the potential of directly boosting the speed of AI algorithms and their ability to improve their performance by analyzing larger sets of data and completing computations at a much faster rate.This could mean that software and applications could become far more efficient, reliable, and powerful with increased scalability.

**8 Conclusion**

1. AI's Impact on Software Development: AI has significantly influenced traditional software development practices by automating code generation, testing, debugging, and deployment processes.This automation has helped to increase the speed of development, reduced oerative errors, and improved quality of generated software.
2. Balancing Human Oversight and AI Autonomy: However, it is crucial to strike a delicate balance between human supervision and the use of artificial intelligence tools.One disadvantage of using AI is that errors or ethical issues can be overlooked if reliance on the technology is excessive.
3. Addressing Ethical and Technical Challenges: Some of the current obstacles include ethical issues, where AI is capable of writing its own code with some level of bias, and technical constraints where integration with other systems, especially outdated ones, may pose a challenge in the employment of AI all throughout the process of the software development life cycle.
4. Future Prospects for AI in Software Development: Moving ahead, the role of AI in the software implementation process will only increase and as technologies like quantum computers come into the software development process, the use of AI is only going to increase.AI is to become a critical component of architecture design as well as a system that will be in charge of creating and maintaining software programs by 2030.